# Eddy current compensated double diffusion encoded (DDE) MRI

Lars Mueller[1]*, Andreas Wetscherek[1], Tristan Anselm Kuder[1], Frederik Bernd Laun[1]

[1]Medical Physics in Radiology, German Cancer Research Center (DKFZ), Heidelberg, Germany

E-mail: muellerl@cardiff.ac.uk

**Abstract**

**Purpose:** Eddy currents might lead to image distortions in diffusion weighted echo planar imaging. A method is proposed to reduce their effects on double diffusion encoding (DDE) MRI experiments and the thereby derived microscopic fractional anisotropy (µFA) .

**Methods:** The twice refocused spin echo scheme was adapted for DDE measurements. To assess the effect of individual diffusion encodings on the image distortions, measurements of a grid of plastic rods in water were performed. The effect of eddy current compensation on µFA measurements was evaluated in the brains of six healthy volunteers.

**Results:** The use of an eddy current compensation reduced the signal variation. As expected, the distortions caused by the second encoding were larger than those of the first encoding entailing a stronger need to compensate for them. For an optimal result, however, both encodings had to be compensated. The artifact reduction strongly improved the measurement of the µFA in ventricles and grey matter by reducing the overestimation. An effect of the compensation on absolute µFA values in white matter was not observed.

**Conclusion:** It is advisable to compensate both encodings in DDE measurements for eddy currents.

**Key words**: microscopic diffusion anisotropy, µFA, double diffusion encoding, eddy current compensation, MRI

**Introduction**

Diffusion weighted imaging (DWI) is a potent tool to reveal structural information in many applications (1,2). One prominent DWI technique is diffusion tensor imaging (DTI), which is commonly used to reveal diffusion and tissue anisotropy by probing the diffusion along different directions (3-5). A popular measure to quantify the anisotropy of the diffusion tensor is the fractional anisotropy (FA). It is usually thought to reflect the diffusion anisotropy in the underlying microscopic structures (e.g. axons), but this correspondence may become ambiguous if the structures are not aligned coherently. Under these circumstances, the microscopic diffusion anisotropy cannot be determined with DTI, but only with advanced DWI techniques, such as those employing multiple successive diffusion encodings (6,7) or a continuous change of diffusion direction (8,9). Among the multiple diffusion encoding techniques, double diffusion encodings (DDE) are used most commonly, where two diffusion encodings along potentially different orientations are applied. The applicability of this technique has been proven in a variety of settings, such as in phantom experiments (10), studies performed ex vivo (11,12), and in vivo in animals (13) and in humans (14-16).

Apart from the time between the diffusion encodings, the mixing time, which may be "short" (6,17,18) or "long" (6,19,20), a second important parameter is the angle between the gradient directions of first and second diffusion encoding (21-23). An arising difficulty is the need to use several different diffusion directions to determine the signal change reliably. Especially in the presence of macroscopic anisotropy, rotationally invariant sampling schemes are needed for this purpose. Schemes have been proposed that use 15 (19) or 72 (24) different direction combinations. The diffusion weighted images, acquired with these direction combinations, must be combined, i.e. added and subtracted, in a specific manner during post-processing to calculate the microscopic anisotropy (19,24). This requires that image distortions, which arise frequently in echo planar imaging (EPI) readouts, are sufficiently small. In particular, varying distortions among the different diffusion direction combinations, which are often caused by eddy currents, are problematic.

Eddy currents are created by fast switching of gradient fields. Modern scanners provide hardware-based eddy current minimization (25), but the long gradient durations and high gradient amplitudes needed for DWI can still produce image artifacts (26). To correct for EPI distortions, the diffusion weighted images can be registered (27) or images with opposing gradient polarity can be used to correct the distortions (28). Another approach is to reduce the effect of eddy currents by sequence design. A prominent approach for single diffusion encoding (SDE) sequences is to use twice refocused spin echo (TRSE) diffusion weightings

(29). While this approach is commonly used in single diffusion encoding experiments, the adaption to double diffusion encoding MRI has not yet been performed to our knowledge. The aim of this work was to perform this adaption and to evaluate its benefits in phantom and in vivo experiments.

**Theory**

To describe the signal in DDE experiments (30-32), the notation of Ref. (24) is used to briefly recapitulate the findings that are important in the present context. The diffusion wave vector is defined as $\boldsymbol{q} = \gamma \, \boldsymbol{G} \delta$, where $\delta$ is the effective gradient duration, $\gamma$ is the gyromagnetic ratio and $\boldsymbol{G}$ the gradient amplitude. The direction of $\boldsymbol{q}$ is defined by the first gradient lobe of the encoding. In the following, it is assumed that $q = |\boldsymbol{q}|$ and the diffusion times $\Delta$ of first and second encoding are identical, so that the two diffusion encodings only differ in their direction (see Fig. 1a).

In the limit of long mixing times $t_\mathrm{m}$ and macroscopically isotropic systems, the signal depends only on the angle $\theta$ between the two wave vectors and not on their absolute direction. In case of macroscopic anisotropy, a dependence on the individual wave vector orientations exists, entailing the need for rotationally invariant direction schemes. One such sampling scheme was proposed (24), which guarantees invariance up to the fourth order of the signal cumulant expansion in $q$, which is the lowest order that yields additional information compared to single encoded experiments (33,34). The sampling scheme of (24) uses 60 direction combinations with perpendicular $\boldsymbol{q}$-vectors and twelve direction combinations with parallel $\boldsymbol{q}$-vectors (with respect of first and second encoding). From images acquired with these 72 direction combinations, the compartment eccentricity $\epsilon$ can be calculated as:

$$\epsilon q^4 = \log\left(\frac{1}{12}\sum S_\parallel\right) - \log\left(\frac{1}{60}\sum S_\perp\right) \quad [1]$$

Here, $\sum S_\parallel$ describes the sum over all parallel measurements and $\sum S_\perp$ the sum over the perpendicular ones. To get rid of the $q$-dependence, one can normalize with a measure of compartment size such as the mean diffusivity (MD). This leads to the following definition of the microscopic fractional anisotropy ($\mu FA$) (8,24,35):

$$\mu\mathrm{FA} = \sqrt{\frac{3}{2}} \sqrt{\frac{\epsilon}{\epsilon + 3/5\,\Delta^2 \mathrm{MD}^2}} \quad [2]$$

The µFA is scaled to range from 0 to 1 and is identical to the fractional anisotropy FA in the case of a single orientation of all compartments under the assumptions that all terms of order $q^6$ and higher can be neglected, that a long mixing time is used, which means the spin displacements during the two diffusion encodings are not correlated and that the exchange between different compartments is negligible during the experiment. If an orientation dispersion is present, the FA decreases, but the µFA remains unchanged.

## Methods

For the eddy current compensation, a model with exponential decay was used (36). The eddy currents are produced by the gradient ramps, which were kept constant. During one encoding, the amplitude of all gradients pulses was kept fixed and only the polarity was changed, so that one can assume an eddy current of amplitude $I_0$ being produced by each ramp. It was assumed that a single eddy current decay time $\tau$ was present. For a standard monopolar diffusion encoding (Fig. 1a shows two monopolar encodings), with gradient duration $\delta$ and diffusion time $\Delta$, this leads to a residual eddy current $I(t)$.

$$I(t) = I_0 \exp\left(-\frac{t}{\tau}\right) \left(\exp\left(-\frac{\delta + \Delta}{\tau}\right) - \exp\left(-\frac{\Delta}{\tau}\right) + \exp\left(-\frac{\delta}{\tau}\right) - 1\right), \qquad [3]$$

where $t = 0$ marks the end of the diffusion encoding. For single diffusion encoding sequences, an eddy current compensated scheme is widely employed that uses an additional refocusing pulse and is therefore called twice refocused spin echo (TRSE) (29). In double diffusion encoding experiments, this scheme can be applied for each encoding (Fig. 1b). Since the directions of the blocks might differ, each encoding must be compensated individually. The decay time $\tau$ was assumed to be 70 ms in this work and the gradient durations were adjusted accordingly to generate eddy current compensated encodings. This time constant was determined in preparatory phantom measurements by varying the assumed decay time (Supporting Figure S1). The minimum of the coefficient of variation was obtained with a time constant of approximately 70 ms. The effective gradient duration is given by:

$$\delta = \delta_1 + \delta_2 = \delta_3 + \delta_4 \qquad [4]$$

All images were acquired with a 3 T MRI (MAGNETOM Prisma, Siemens Healthcare, Erlangen, Germany). Its maximal gradient amplitude is 80 mT/m, with a slew rate of 200 T/m/s.

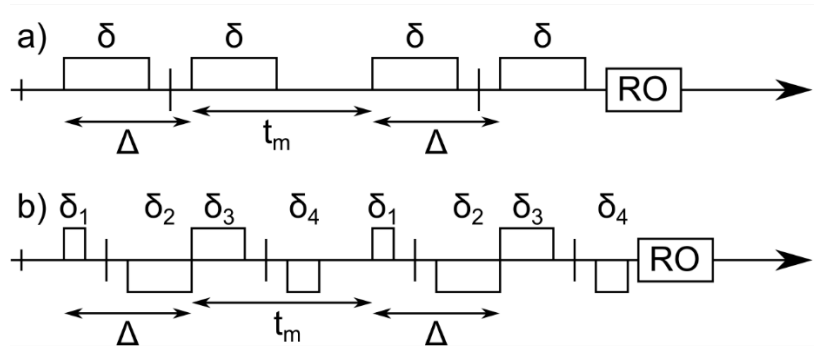

**Figure 1:** Sequence schemes used for DDE. a) Sequence version with no eddy current compensation. b) Sequence scheme where both encodings are eddy current compensated.

The unwanted echo pathways were suppressed using pairs of crusher gradients around each refocusing pulse. These gradients were chosen to be perpendicular to their respective diffusion encoding as described in (37). In the case of compensated encodings, the crusher pairs of each single encoding were also orthogonal to each other. In the case of parallel crusher gradients in the first and second encoding, the ones in the second encoding were rotated by 45° around the second diffusion encoding. The zeroth gradient moment was equal for all crusher gradient pulses and ensured a dephasing of $8\pi$ in one voxel. In preparatory experiments, it was ensured that crusher gradients would not afflict the evaluation of the effect of eddy currents by measuring once with doubled zeroth gradient moment of the crusher gradients and once with a scheme with fixed gradient directions of the crusher gradients using a liquid phantom.

Two kinds of experiments were performed. Phantom measurements were executed to quantify the actual effect of eddy currents. Four different combinations were used for first and second encoding (compensated=c, uncompensated=u): u/u, u/c, c/u, and c/c. Head measurements of six healthy volunteers were performed with the combinations u/u and c/c to evaluate the importance in applications.

The phantom consisted of an equidistant grid of plastic rods in an octagonal container filled with water. The total height was 330 mm with a thickness of 100 mm and the distance between adjacent rods was 2 cm. All direction combinations of plus and minus in read, phase and slice-direction, which coincided with x, y and z-axis in the scanner system, were used for the diffusion gradients, leading to a total of 36 direction combinations.

The compensated encodings were each performed with $\delta$ = 10.87 ($\delta_1$ = 3.14, $\delta_2$ = 7.73, $\delta_3$ = 8.14, $\delta_4$ = 2.73) ms, $\Delta$ = 19 ms and G = 78.7 mT/m, leading to $b$ = 500 s/mm², $q$ = 228 mm$^{-1}$. For the uncompensated encoding, these parameters were $\delta$ = 9.44 ms, $\Delta$ = 17.56 ms, G = 77.4 mT/m, b=500 s/mm² and q = 196 mm$^{-1}$. The time for each slice selection with their corresponding crusher gradients was 5.49 ms.

The echo times were 94, 105, 85, 96 ms with mixing times of 29.44, 19, 19.44 and 19 ms for u/u, c/u, u/c and c/c, respectively. Other imaging parameters were TR = 3000 ms, FoV = 380 × 380 mm², nominal resolution 3.8 × 3.8 mm², slice thickness of 5 mm and 10 slices. The measurements were performed with the body coil.

For comparison of image artifacts, the coefficient of variation, which is the ratio of the standard deviation and the mean of the signal obtained with the 36 direction combinations, was calculated. For a baseline estimate, one identical measurement was performed with 36 repetitions and both encodings pointing along read direction.

For the in vivo experiments, brain images of 6 healthy volunteers were acquired. The 72 direction combinations were taken from (24) (see Theory section) and µFA was determined voxelwise. In the compensated version, the gradient durations were $\delta_1$ = 2.34 ms, $\delta_2$ = 8.67 ms, $\delta_3$ = 8.84 ms, $\delta_4$ = 2.17 ms, with $\Delta$ = 19.5 ms, $t_m$ = 29.5 ms, G = 79 mT/m and TE = 104 ms, in the uncompensated encodings $\delta$ was 9.26 ms, $\Delta$ = 17.74 ms, $t_m$ = 25.26 ms, G = 78.3 mT/m and TE = 86 ms. The $b$-value of each encoding was always 500 s/mm², leading to $q$-values of 193 and 233 mm$^{-1}$ for the uncompensated and compensated encodings, respectively. The other imaging parameters were identical in both cases: TR = 4000 ms, FoV = 300 × 270 mm² with a nominal in-plane resolution of 3 × 3 mm² and a slice thickness of 3 mm, 20 slices were acquired. Additionally, a partial Fourier factor of 6/8 was used with a GRAPPA factor of 2. The time for each slice selection was 5.85 ms. A 64-channel head coil was used in all volunteer measurements. For determining MD, an additional image without diffusion weighting was acquired. MD was then taken as one third of the trace of the diffusion tensor $\underline{\underline{D}}$ obtained by performing a fit of equation 4 in (38) to all measurements:

$$S(\boldsymbol{q}_1, \boldsymbol{q}_2) = S_0 \exp\left(-\frac{1}{2}\boldsymbol{q}_1^T \underline{\underline{D}}\, \boldsymbol{q}_1 \Delta - \frac{1}{2}\boldsymbol{q}_2^T \underline{\underline{D}}\, \boldsymbol{q}_2 \Delta + \boldsymbol{q}_1^T \underline{\underline{Q}}\, \boldsymbol{q}_2\right), \qquad [5]$$

with the correlation tensor $\underline{\underline{Q}}$, giving the strength of correlation between both diffusion encodings. This equation had been derived by a cumulant expansion of the signal assuming small $q_1$ and $q_2$. Here, $\underline{\underline{Q}}$ was used to capture the residual correlation, since the long time limit might not be reached. All parameter maps were calculated voxelwise using the magnitude images from the scanner. Additionally, interpolated parameter maps were calculated using the Matlab function "interp3". The interpolated FA maps were also used for the definiton of regions of interest (ROI) in a high FA fibre region and in a fibre crossing with low FA.

The brain was segmented by means of absolute values of MD and FA. Voxels with a MD between 0.35 and 1 µm²/ms and a FA higher than 0.35 were marked as white matter, while voxels with MD between 0.5 and 1.3 µm²/ms and a FA between 0.02 and 0.15 were considered to be grey matter (39). Voxels with MD higher than 1.5 µm²/ms were classified as ventricles.

## Results

**Phantom Data**

The phantom measurements show an overall reduction in signal variation when using an eddy current compensation. This can be seen in the maps of the coefficient of variation of the phantom (Fig. 2). At the edges perpendicular to the phase encoding direction (top-down in Fig. 2b-e), the coefficient of variation increases drastically, which indicates image distortions along this direction. The areas with a higher coefficient of variation decrease from no eddy current compensation (u/u) over c/u and u/c to a full eddy current compensation (c/c). Inside the phantom, the coefficient of variation shows the grid structure, where the single grid points are blurred in phase direction. Again, the coefficient of variation visibly is minimized for c/c.

Fig. 2f shows the mean coefficient of variation in the white regions of interest (ROI) depicted in Fig. 2b-e for all ten slices. The benefit of eddy current compensation is most prominent for the second encoding, while the first encoding has much smaller influence on distortion artifacts. The level of the baseline measurements cannot be achieved by any scheme.

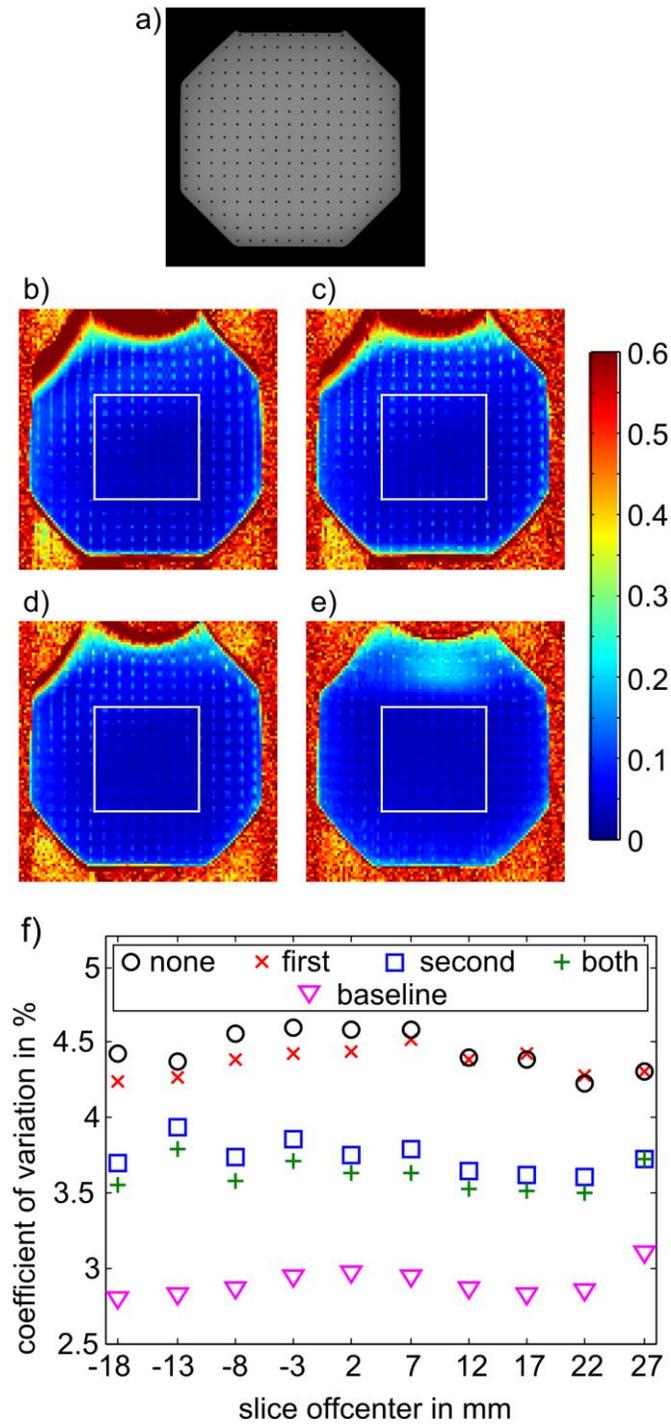

**Figure 2:** Maps of coefficient of variation over 36 diffusion directions. a) High resolution image of the phantom. The edges of the phantom and the individual rods show a clear reduction of image artifacts from u/u (b) over c/u (c) and u/c (d) to c/c (e). The white square (41² pixels) was used for calculating the mean value (f). The heightened coefficient of variation above the ROI in the c/c (e) measurement is due to concomitant field artifacts.

**Brain Data**

In Fig. 3, FA maps, and µFA maps for the fully uncompensated (u/u) and fully compensated (c/c) acquisition are shown for an exemplary slice of each volunteer. The µFA shows a wider area of high values in both cases in comparison to the FA. A main difference between the uncompensated and compensated µFA maps can be observed in the ventricles, where one would expect a low µFA. In the uncompensated µFA maps, the ventricles appear bright (arrows), an effect which is clearly reduced in the compensated maps. A figure with enlarged central areas is provided as Supporting Figure S2.

The mean values of the µFA in white and grey matter and in the ventricles are stated in Tab. 1 (40). A clear reduction of the µFA in the ventricles is observed when using the compensated scheme instead of the uncompensated scheme. Additionally, a reduction of the µFA in grey matter is visible, while the µFA in white matter changes only slightly.

In white matter fiber crossings (Fig. 4), the FA is reduced (Fig. 4c, green arrow). The FA in the upper ROI (Fig. 4b) in the crossing is 0.35 ± 0.03, while it is 0.64 ± 0.04 inside the fiber (lower ROI) for the volunteer data shown. This reduction almost vanishes in µFA maps. The uncompensated µFA in the fiber is 0.89 ± 0.02 and in the crossing it is 0.86 ± 0.03. For the compensated version it takes the values 0.88 ± 0.02 in the fiber and 0.87 ± 0.03 in the crossing. Fig. 4 displays only minor differences in the µFA between uncompensated and compensated acquisition, but the ventricle (Fig. 4d-e, red arrow) appears bright in the µFA map acquired with the u/u scheme, which is not seen in the µFA map acquired with the c/c scheme.

**Table 1:** Mean ± standard deviation of the µFA in different brain regions

|  | white matter | | grey matter | | ventricles | |
|---|---|---|---|---|---|---|
| volunteer | u/u | c/c | u/u | c/c | u/u | c/c |
| 1 | 0.79 ± 0.13 | 0.75 ± 0.16 | 0.54 ± 0.17 | 0.51 ± 0.19 | 0.47 ± 0.15 | 0.18 ± 0.23 |
| 2 | 0.78 ± 0.14 | 0.79 ± 0.14 | 0.57 ± 0.17 | 0.49 ± 0.19 | 0.36 ± 0.15 | 0.14 ± 0.21 |
| 3 | 0.79 ± 0.13 | 0.77 ± 0.14 | 0.52 ± 0.18 | 0.45 ± 0.21 | 0.53 ± 0.23 | 0.33 ± 0.34 |
| 4 | 0.74 ± 0.16 | 0.75 ± 0.16 | 0.57 ± 0.20 | 0.53 ± 0.18 | 0.61 ± 0.20 | 0.18 ± 0.23 |
| 5 | 0.76 ± 0.16 | 0.76 ± 0.15 | 0.63 ± 0.19 | 0.51 ± 0.20 | 0.59 ± 0.23 | 0.29 ± 0.28 |
| 6 | 0.72 ± 0.17 | 0.65 ± 0.21 | 0.60 ± 0.20 | 0.53 ± 0.21 | 0.68 ± 0.18 | 0.58 ± 0.31 |

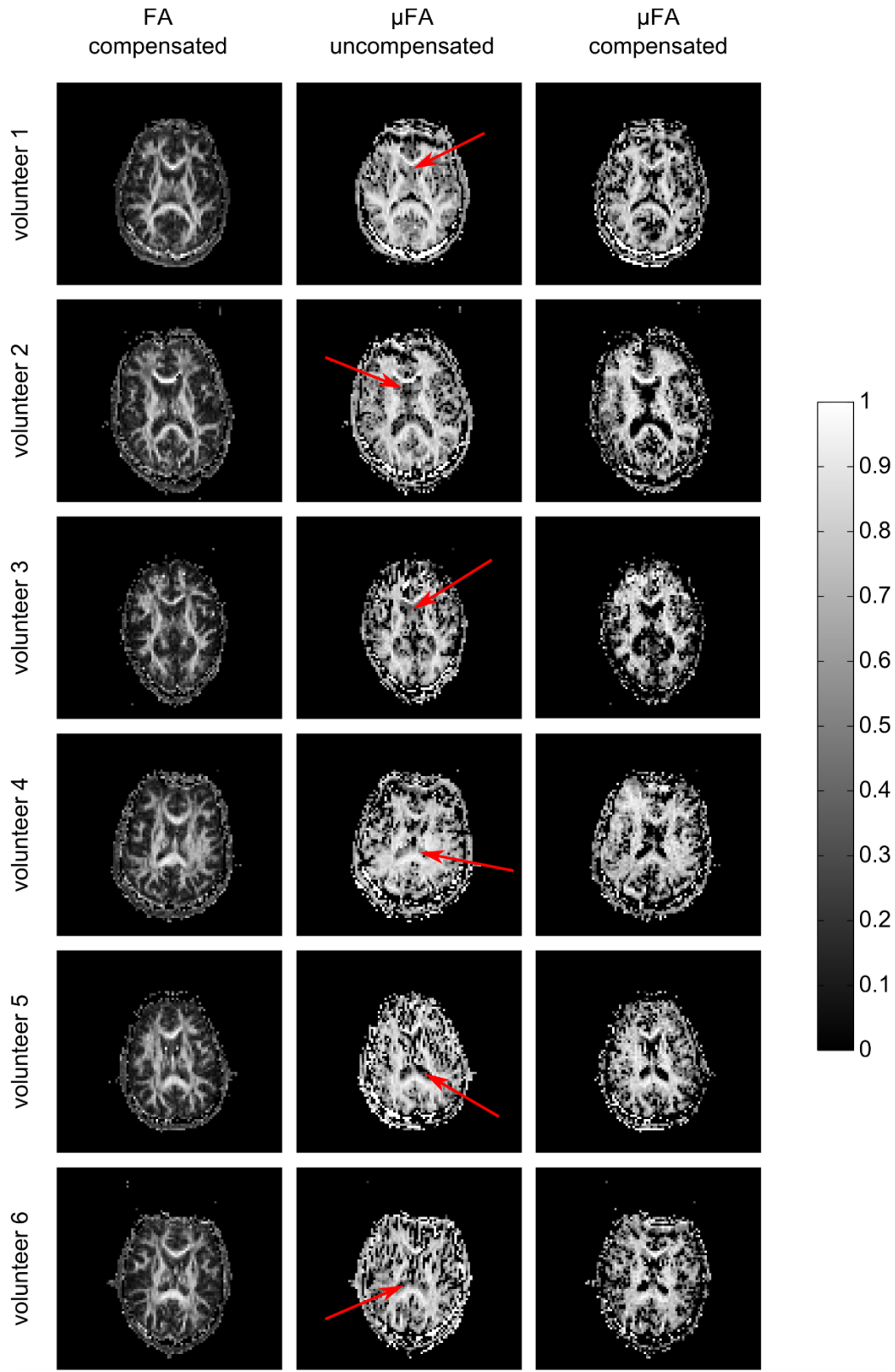

**Figure 3:** Maps of the FA and \mu FA of the six volunteers in a representative slice. The µFA shows larger areas of high anisotropy. The eddy current compensated maps show less artifacts, especially in the ventricles.

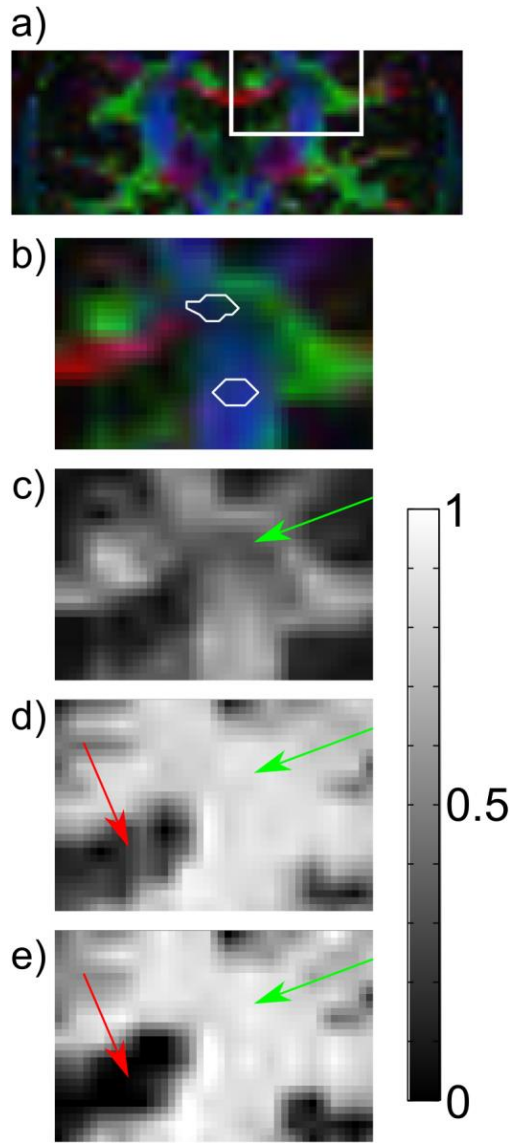

**Figure 4:** Coronal slice for 1 volunteer. In fiber crossings, the FA is reduced (a-c), where a) shows the color encoded FA of the complete slice and the area, which is magnified in (b-e). The reduction of FA (green arrow) in the crossing compared to single fiber regions is not present in the µFA maps (d-e). The uncompensated measurement (d) shows visibly nonzero µFA values in the ventricles (red arrow), which disappears in the compensated measurement (e). The images were tricubically interpolated to the threefold resolution.

## Discussion

In this work, the use of an adjusted twice refocused spin echo gradient scheme was shown to be a viable approach to reduce eddy current artifacts in DDE. The use of the proposed eddy current compensation reduces the need for other distortion corrections that might require additional images (15,28) and/or use post-processing algorithms for the correction of the images, e.g. by registration (27,41). These approaches require sophisticated methods, because image intensity changes significantly with the diffusion direction (42) and they have to deal with the low signal-to-noise ratio in DWI.

An improvement to the proposed compensation might be to assume a biexponential eddy current decay and to compensate the encodings for different decay times, which was shown to be beneficial in the context of the single diffusion encoding TRSE experiments (43). Our phantom measurements revealed a clear difference between the baseline measurement and the c/c scheme in the coefficient of variation, which might be decreased by such an approach.

The eddy current compensation used here could also be applied in filter exchange imaging (39,44), especially when investigating a directional dependence of the exchange measurement (45). In the estimation of pore sizes (17,23,46,47) the zero mixing time limit is used. These methods also rely on the signal variation in dependence of the angle between the encodings and could profit from eddy current compensation. The proposed approach does not compensate for concomitant fields, whose influence could be reduced by appropriate adjustments to gradient amplitudes and durations (48) or by an adjusted image reconstruction (49). In the head measurements, the concomitant fields could be observed by a phase modulation (data not shown), but the phase variation was well below $2\pi$ in a voxel, thus not resulting in signal dropouts. Although phase modulations in slice direction could not be detected by this approach, they can be assumed to be of the same order as in-plane phase modulations and therefore should not lead to any significant magnitude changes.

The improvement in the μFA measurements in this work mainly stems from the reduction of falsely heightened microscopic anisotropies, for example, in the ventricles and grey matter. Another source of such artifacts might be background gradients, for example arising from different susceptibilities in adjacent tissues (50). These background gradients can be compensated in SDE using bipolar gradients in a STEAM based sequence (51) as well as using a spin echo based sequence (52). It is possible to adapt both approaches to DDE (47,50).

The determination of microscopic anisotropy with DDE was demonstrated in vivo measuring the microscopic anisotropy MA. It scales in the range between 0 and 1, but is not identical to the µFA (15). Values of MA in different white matter ROIs were reported to lie between 0.629 ± 0.046 and 0.947 ± 0.017, while most white matter voxels exhibit a MA between 0.8 and 0.9 (15).

Another way to determine the microscopic anisotropy is via q-vector magic angle spinning (qMAS) (9,53). This method assumes Gaussian diffusion in the pores with a time-independent diffusion coefficient and a negligible exchange during the diffusion encoding(54). If these assumptions are met, qMAS is assumed to yield the same µFA as the DDE method used here (54). The reported µFA values for qMAS in several white matter tracts range from 0.93 ± 0.01 to 1.02 ± 0.02 (53) and are thus considerably higher than herein presented values. One reason could be the approximately 3 times higher diffusion time in (53), allowing the spins to probe the anisotropy better.

One has also to consider that the assumptions made for calculating the µFA might not have been met. The mixing time was not long enough to lose all correlations between the two encodings, which might lead to systematic errors (20,31,55) owing to a residual short mixing time effect. This could be corrected for by including the 12 antiparallel encodings (15) additionally to the 12 parallel ones. Since the measurement parameters of both sequences were quite similar, the here observed improvement should not be due to a reduction of these systematic errors, nor to the slightly different values of $\delta$ and $\Delta$. Concerning the magnitude of the q-values, Eq. 2 can only be applied if the influence of higher order terms, i.e. terms of order $q^6$, or $b^3$, is negligible. In general, it is difficult to assess the effect of higher order terms, because their size is not known a priori. In comparison to previous studies, the q-values used here (193 and 233 mm$^{-1}$) are somewhat higher than in two studies in the human brain ($\approx$ 160 mm$^{-1}$) (15,16), but still lower than in ex vivo experiments on fixed vervet monkey brains (560 mm$^{-1}$) (24). Thinking in terms of b-values, Jensen and Helpern estimate that a maximal b-value of 2000 s/mm² leads to estimates of the kurtosis (which is of order $q^4$ or $b^2$) with an accuracy of roughly 20% (56). The here used experiments can roughly be treated on the same footing and thus it is likely that the here used q- and b-values are reasonably small, such that higher order terms could be neglected.

The extra refocusing pulses used in our work lead to a reduction of the signal of about 20% in the phantom experiments. Additionally, the echo time might be prolonged and longer repetition times might be needed, due to the higher energy deposition. But still the advantage of the reduced eddy current distortions outweighs the signal loss, since the white matter showed comparable µFA values in u/u and c/c measurements and at the same time falsely heightened values were reduced in the c/c scheme in the other brain areas. Therefore, in conclusion, it is advisable to use a twice refocusing scheme for the diffusion encodings in DDE applications.

**Acknowledgement**

Financial support by the DFG (grant no. KU 3362/1-1 and LA 2804/2-1) is gratefully acknowledged.

**Supporting Material**

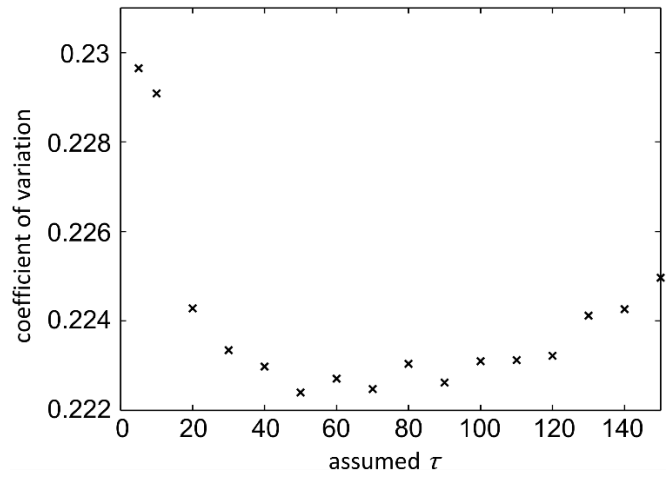

**Supporting Figure S1**: Determination of used eddy current decay time. For this measurement the total duration of the diffusion encodings were kept fixed and the assumed decay time $\tau$ was varied. Other measurement parameters were identical to the phantom experiments. The coefficient of variation over the whole slice of the grid phantom shows a wide minimum around 70 ms.

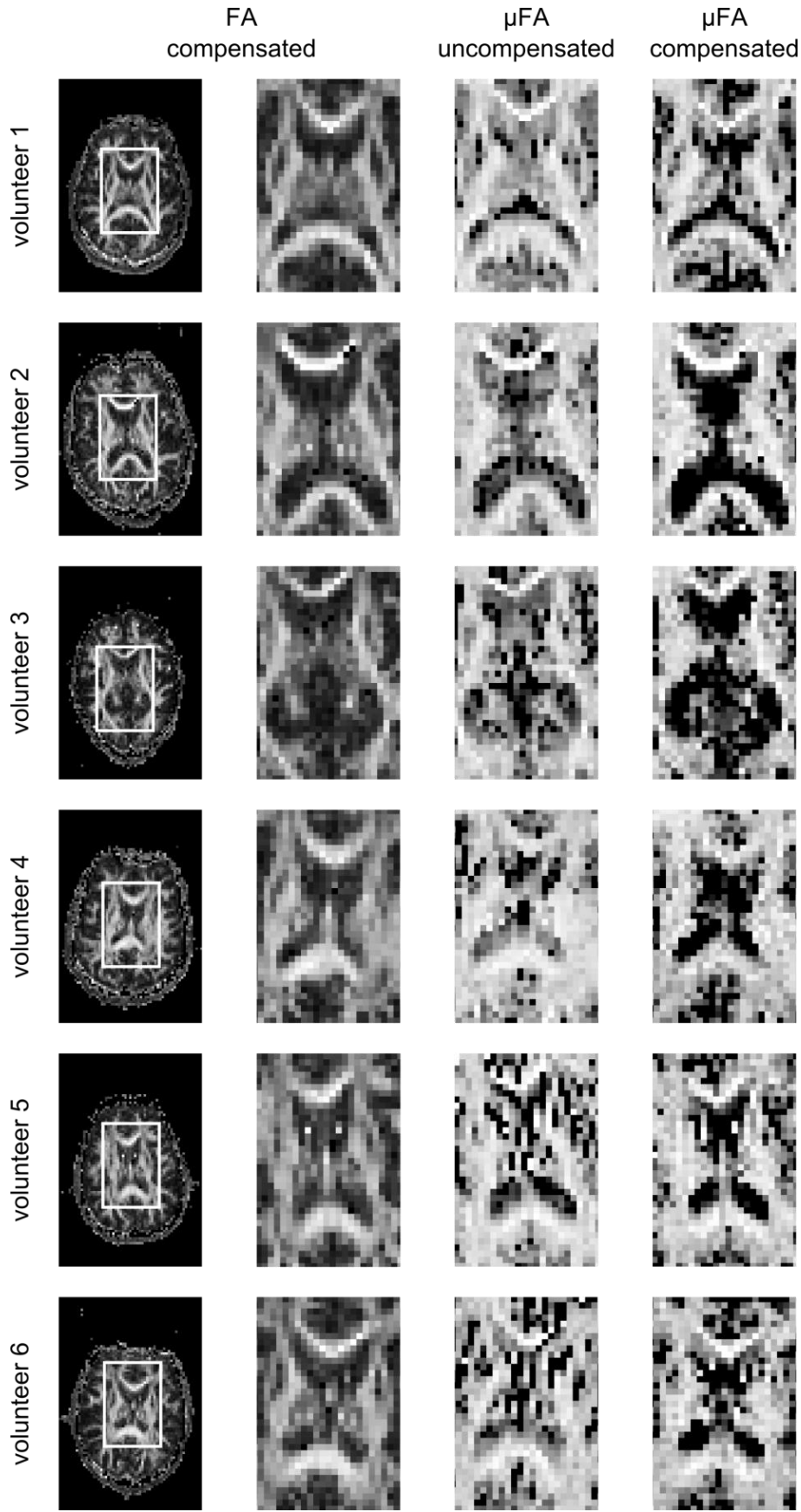

**Supporting Figure S2**: Enlarged central area of Fig. 3. These are the same slices as in Fig. 3, but show only the enlarged central area. The falsely heightened µFA in the ventricles is clearly visible in the uncompensated measurements. The images were not interpolated.

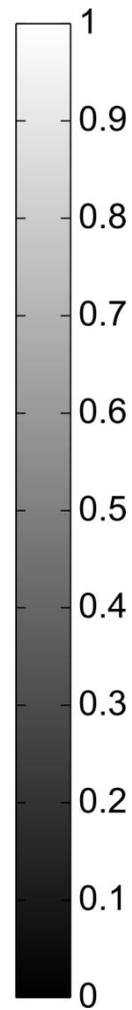